\documentclass{elsart}
\usepackage{amssymb}
\usepackage{epsfig,epsf,graphicx}
\usepackage{amsmath}

\newcommand{\be}{\begin{equation}}
\newcommand{\ee}{\end{equation}}

\newcommand{\bee}{\begin{eqnarray}}
\newcommand{\eee}{\end{eqnarray}}

\def\be{\begin{eqnarray} &&}

\def\ee{\end{eqnarray}}
\def\bew{\begin{widetext}}
\def\ew{\end{widetext}}

\usepackage{graphicx}
\usepackage{dcolumn}
\usepackage{bm}

\usepackage{color}
\newcommand{\mulato}{\textit{WIMG }}
\newcommand{\cmulato}{\textit{WIMG}, }
\newcommand{\pmulato}{\textit{WIMG}. }

\begin{document}
\begin{frontmatter}
\title{A Democratic Gauge Model for Dark/Visible Matter Symmetry}

\author{O. Oliveira$^{*,\dagger}$, C. A. Bertulani$^{\ddagger}$, M. S. Hussein$^\parallel$,}
\author{W. de Paula$^*$ and T. Frederico$^*$}

 \address{$^\dagger$Departamento de F\'{\i}sica, Universidade de Coimbra, 3004-516 Coimbra, Portugal}
\address{$^\ddagger$ Department of Physics and Astronomy, Texas A\&M University-Commerce,
                                   Commerce TX 75429, USA}
\address{$^\parallel$ Instituto de F\'{\i}sica, Universidade de S\~ao Paulo, Caixa Postal 66318,
                                     05314-970 S\~ao Paulo, SP, Brazil}
 \address{$^*$Departamento de F\'{\i}sica, Instituto Tecnol\'ogico de Aeron\'autica, DCTA
                               12.228-900, S\~ao Jos\'e dos Campos, SP, Brazil}

\date{\today}
\maketitle

\begin{abstract}
We develop a model for visible matter-dark matter interaction based on the
exchange of a weakly interacting massive gauge boson called herein the WIMG.
Our model hinges on the assumption that all known particles in the visible matter
have their counterparts in the dark matter. We postulate six families of particles five of which are dark.
This leads to the unavoidable postulation of six parallel worlds, the visible one
and five invisible worlds. We give arguments on particle decays and lifetimes that
set a limit on the mass of the WIMG, the gray boson responsible for the very meager
communication among these worlds. The 5:1 ratio of dark to visible matter is taken for granted.

\end{abstract}


\begin{keyword}
Dark Matter, Weakly Interacting Massive Particles, Parallel Universes
\end{keyword}
\end{frontmatter}


It is widely accepted that our Universe is composed of 70\% Dark Energy (DE), 25\% Dark Matter (DM) and
only about 5\% Visible Matter (VM). The existence of the DM and the DE is inferred to through observation
made on the VM (galaxies, black holes) or, in the latter case, to cosmological consequences of having a
negative pressure in the equation of sate (the cosmological constant problem). The interaction between
DM and DE,  and DM and VM have received a great deal of attention over the last decades 
(see e.g. \cite{bertone05,bertone10,He2011} and references therein). In particular
the latter interaction has evoked new scenarios that go beyond the Standard Model, such as the existence of
Weakly Interacting Massive Particles (WIMPs), Sterile Neutrinos, Axions,  etc. The discovery of any of these
will constitute a most important input to our understanding of the nature of DM.
Experimental limits on the interaction and masses of WIMPs inferred to through recoil measurement of nuclear
targets are now available \cite{CDMS,XENON100,XENON100-2011,DAMA,DAMA2010,DAMA2010b,COGENT}.
More work is required though, to better our understanding of
how the VM particles interact with the DM counterparts \cite{cui11,Hamed2009}. 

The ratio of DM to VM inspired us to pursue a new venue which 
lead to the present work. We develop a model which assumes that for each charge family of particles
in the VM there are a corresponding 5 charge families of Dark particles in the 5 parallel universes of the DM.
Namely, one of them is our visible universe which is composed of two charge families of quarks, u, c, and t of
charge 2e/3; d, s and b, of charge -e/3; the leptons and their corresponding neutrinos. The other five
universes contain only DM in  similar two charge families of dark quarks, $u_d, c_d, t_d$ of charge
2e/3, $d_d, s_d, b_d$, of charge -e/3; leptons, $e_d$, $\mu_d$ and $\tau_d$, and dark neutrinos.
Dark photons can only interact with the Dark charges. The weakly interacting massive gauge
boson responsible for the new force between these parallel universes is baptized as the \pmulato
The current Letter describes the above six-parallel-universe picture of the interaction
between DM and VM. Several consequences of the possible existence of the \mulato boson are
looked at through calculation of the lepton anomalous magnetic moments (which sets a bound on
the \mulato mass of several hundreds GeVs), lepton flavor violation processes, $\mu$, and $\beta$
decays of D and B mesons, and the SM tree level forbidden process  $e^{+} + p \rightarrow \mu^{+} + \Lambda$,
or $\Lambda_c$, allowed through a \mulato exchange in our model.  In the following we give a detailed account
of our DM-VM 6 parallel Universes model.


The gauge theory includes the spin-1 \mulato field $M^a_\mu$, the matter fields $Q_f$, where $f$ is a flavor index,
and a scalar field $\phi^a$ belonging to the adjoint representation of $SU(3)$ color group. As in the Standard Model,
the scalar field is required to provide a mass to $M^a_\mu$ to ensure that the \mulato interaction is of short
distance. In this way, new contributions to the long distance nature of the gravitational force are avoided.

In what concerns the matter fields, the visible multiplets are
\begin{equation}
  Q_1 = \left( \begin{array}{c} u \\ c \\ t \end{array} \right) , \quad
  Q_2 = \left( \begin{array}{c} d \\ s \\ b \end{array} \right) , \quad
  Q_3 = \left( \begin{array}{c} e \\ \mu \\ \tau \end{array} \right) , \quad
  Q_4 = \left( \begin{array}{c} \nu_e \\ \nu_\mu \\ \nu_\tau \end{array} \right) \, .
  \label{matter}
\end{equation}
It will be assumed that the $Q_f$'s belong to the fundamental representation of SU(3),
with all members of each multiplet having the same electrical charge.
Further, in the following we will consider two different types of multiplets: (i) the matter fields in $Q_f$
do not include a chiral projector, called non-chiral theory below; (ii) the fields in $Q_f$ are all left-handed,
called chiral theory below, and a $\gamma_L = ( 1 - \gamma_5 )/2$ should be attached to each field in (\ref{matter}).

In order to comply with the relative abundance between dark and visible matter, we postulate that for each
visible multiplet $Q_f, f = 1, \dots, 4$, there are five dark matter multiplets. The dark matter multiplets are
built just as the visible multiplets. However, given that the dark matter does not seem to couple with the
electromagnetic field, it will be assumed that the dark multiplets all have zero electrical charge.
As discussed below, within each multiplet the interaction is invariant with respect to a U(1) transformation.
Therefore, we can define a dark-photon which couples only with the dark multiplets. The matter multiplets have
different U(1) charges, an electric charge, which vanishes for the dark multiplets, and a dark electric charge,
which vanishes for the visible matter. Then, from the point of view of the theory, visible and dark matter are
treated democratically, i.e. apart from the relative abundance of the two different types of matter, they are
distinguished only by their electric and dark electric charge. Further, it will be assumed that dark and visible
matter can only interact via \mulato exchange. In order to explain the non-observation of such type interactions,
the \mulato field must be a massive field with a mass much larger than the electroweak mass scale.

The Lagrangian  for the gauge theory reads
\begin{eqnarray}
 \mathcal{L}  =  - \frac{1}{4} F^a_{\mu\nu} F^{a \, \mu\nu} ~ + ~
 \sum_f \overline Q_f \left\{ i \gamma^\mu D_\mu - m_f \right\} Q_f
 ~ + ~ \nonumber \\ +\frac{1}{2} \left( D^\mu \phi^a \right) \left( D_\mu \phi^a \right) - V_{oct}( \phi^a \phi^a )
 ~ + ~ \mathcal{L}_{GF} ~ + ~ \mathcal{L}_{gh}
 \label{lagrangeano}
\end{eqnarray}
where $D_\mu = \partial_\mu + i g_M T^a M^a_\mu$  is the covariant derivative,
$T^a$ stands for the generators of $SU(3)$ color group, $m_f$ the current quark mass matrix and
$V_{oct}$ is the potential energy associated with $\phi^a$. A sum over the six species of matter, VM and DM,
is implicit in our notation.
$\mathcal{L}_{GF}$ is the gauge fixing part of the Lagrangian and $\mathcal{L}_{gh}$ contains the ghost terms.
The various terms in $\mathcal{L}$ are gauge invariant, with the exception of $\mathcal{L}_{GF}$ and
$\mathcal{L}_{gh}$. However, for example in the Landau gauge, $\mathcal{L}_{GF} + \mathcal{L}_{gh}$ is
BRST invariant. The Lagrangian density (\ref{lagrangeano}) fixes unambiguously the interactions between
visible and dark matter.

The Lagrangian density $\mathcal{L}$ has several symmetries besides the local SU(3) gauge invariance.
It is invariant under U(1) gauge transformations within each multiplet. This allows the introduction of
multiple photon-like fields. Apart from the mass matrices in $\mathcal{L}$, the Lagrangian is invariant
under flavor transformations. Further, setting $g_M = 0$, the Lagrangian is invariant under global SU(3)
transformations within each multiplet $Q_f$. This freedom, allows for the introduction of unitary matrices
associated with each multiplet to diagonalize the mass matrices that, although of different origin, play
a similar role as the CKM-matrix in the Standard Model Lagrangian.
In this way, the model can accommodate neutrino mixing. Further, if one takes the
democratic principle seriously, one can build a standard model like Lagrangian for the dark sector.
Despite its rich structure, in the present work we will not explore the
features just mentioned in this paragraph. Here, we are mainly concerned with the mass scales and
phenomenological implications of the theory summarized in (\ref{lagrangeano}).
The implications of all the symmetries of $\mathcal{L}$ will be the subject of a future publication.

The \mulato should not change the long distance properties of the gravitational interaction. The only way this can
be achieved is if the \mulato is a massive particle. In order to generate a mass to $M^a_\mu$ keeping gauge
invariance, one has to rely on scalar fields. According to the Goldstone theorem \cite{Huang1992}, the
Higgs mechanism leaves a number of components of $M^a_\mu$ massless and, to keep the long distance forces
unchanged, the Higgs mechanism must be excluded as a way to give mass to the gauge fields. In \cite{Oliveira2011},
the authors propose a mechanism for mass generation via the introduction of a scalar condensate which complies
with gauge invariance. Further, the mass generation mechanism provides the same mass for all the components of
the gauge field. In the following, we will assume that the \mulato acquires mass through this mechanism, which
we are about to describe.

The kinetic term associated with the scalar field accommodates a mass term for the \mulato field. The gauge field
mass term is associated with the operator
\begin{equation}
  \frac{1}{2} \, g^2_M \, \phi^c (T^a T^b)_{cd} \phi^d M^a_\mu M^{b \, \mu} \, .
\end{equation}
The scalar field cannot acquire a vacuum expectation value without breaking gauge invariance. However,
to generate a mass for the \cmulato it is sufficient to assume a non-vanishing boson condensate
$\langle \phi ^a \phi^b \rangle$. The origin of this condensate can be associated with local fluctuations of
the scalar field. From now on, we assume that the dynamics of the scalar field is such that
\begin{equation}
  \langle \phi ^a \rangle = 0 \qquad \mbox{ and } \qquad \langle \phi^a \phi^b \rangle = v^2 \delta^{ab} \, .
\end{equation}
Given that for the adjoint representation $\mbox{tr} \left(  T^a T^b \right)= 3 \, \delta^{ab}$,
the square of the \mulato mass reads
\begin{equation}
  M^2  = 3 \, g^2_M   v^2 \, .
  \label{massa_gluao}
\end{equation}
Note that the condensate $\langle \phi^a \phi^b \rangle$, i.e. $v^2$, and therefore the \mulato mass
is gauge invariant. The proof of gauge invariance follows directly from the transformations properties
of $\phi^a$. We have assumed that the real scalar field $\phi$ belongs to the adjoint representation of SU(3).
However, the same mechanism can be applied if $\phi$ belongs to the fundamental representation of the gauge group.
The main difference being that for the adjoint representation $\phi$ is real and, therefore, has zero charge,
while for the fundamental representation $\phi$ is complex field and, in this way, couples to the photon or
dark-photon fields.

\begin{figure}[h]
\centering
\epsfig{file=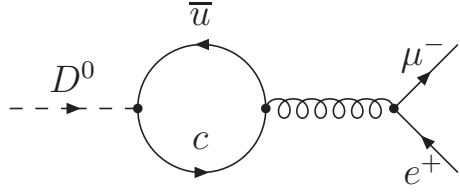,angle=0, width=6cm}
\vspace{.5cm}
\caption{$D^0$ decay to $(\mu^+,e^-)$ by \mulato exchange process.} \label{ddecay}
\end{figure}
\begin{figure}[h]
\centering
\epsfig{file=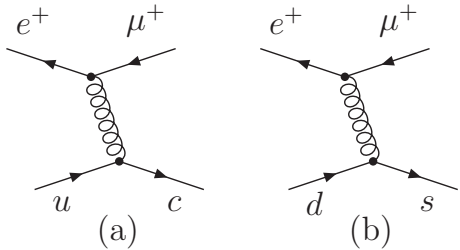,angle=0, width=6cm}
\vspace{.5cm}
\caption{Lepton-quark scattering with violation of the lepton family number and flavor exchange by \mulato
mediated processes.
Positron conversion to antimuon and flavor exchange $u\to c$ (a) and $d \to s$ (b). } \label{flavorx}
\end{figure}

We now aim to discuss the phenomenology associated with the new interaction. For the first hadronic visible family, the
\mulato - quark interaction part of the Lagrangian is
\begin{eqnarray}
 \mathcal{L}_{Mq} =
  \frac{g_M}{2}Ê\left[ \overline u \, \gamma^\mu \, c \right] \left( M^1_\mu - i M^2_\mu \right) +
  \frac{g_M}{2}Ê\left[ \overline u \, \gamma^\mu \, t \right]  \left( M^4_\mu - i M^5_\mu \right) +
  \nonumber\\ + \frac{g_M}{2}Ê\left[ \overline c \, \gamma^\mu \, t \right] \left( M^6_\mu - i M^7_\mu \right) +
  \frac{g_M}{2}Ê\left[ \overline u \, \gamma^\mu \, u \right] \left( M^3_\mu + \frac{1}{\sqrt{3}} M^8_\mu \right) +
\nonumber \\ +
  \frac{g_M}{2}Ê\left[ \overline c \, \gamma^\mu \, c \right] \left( - M^3_\mu + \frac{1}{\sqrt{3}} M^8_\mu \right) +
  \frac{g_M}{2}Ê\left[ \overline t \, \gamma^\mu \, t \right] \left( - \frac{2}{\sqrt{3}} M^8_\mu \right)  +  h.c.
  \label{L_q_mulato}
\end{eqnarray}
The remaining families having similar types of interactions. The new vertices can give rise to flavor changing
type of processes but only if the flavor changing occurs only within the same family. Given that the \mulato
has no electrical charge, it seems that it can give rise to flavor changing neutral processes which are, at most,
suppressed by $\sim g^2_M/M^2$. However, the flavor structure of (\ref{L_q_mulato}) and given that
the \mulato propagator is flavor diagonal, the $S-$matrix element for these processes vanishes.
For example, as discussed below, the \mulato vertices give no contributions to the lepton flavor violation
processes reported at the particle data book \cite{pdg2010}. In this sense, the \mulato interaction is
compatible with the GIM (Glashow-Iliopoulos-Maiani) mechanism of the standard model and the flavor changing
neutral currents should remain suppressed at high energies. However, the interaction Lagrangian (\ref{L_q_mulato})
allows lepton family number violation, as we will discuss (see e.g. figures
\ref{ddecay} and \ref{flavorx}).


\begin{figure}[h]
\centering
\epsfig{file=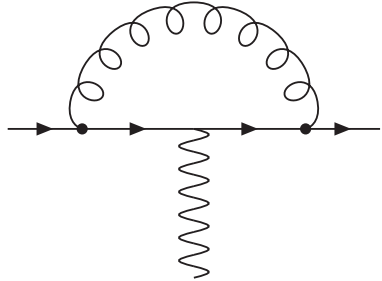,angle=0, width=5cm}
\caption{Lepton-photon vertex correction by \mulato exchange.} \label{photonvertex}
\end{figure}

The new gauge boson interactions provide corrections to the lepton-photon vertex which
contribute to the lepton anomalous magnetic moment as depicted in fig.\ref{photonvertex}.
The new contributions to $(g-2)/2$ coming from the \mulato are UV-finite,
 reading
\begin{equation}
 a_{e,\mu} =  \frac{g^2_M}{16 \pi^2} \left( \frac{m_{e,\mu}}{M} \right)^2\left(\frac53-
\frac34\frac{m_\tau}{m_{e,\mu}}\right) \ \ {\mbox {  and  }} \ \
 a_\tau =  \frac{7\, g^2_M}{96 \pi^2} \left( \frac{m_\tau}{M} \right)^2 \ ,
\label{eq:g2_nonchiral}
\end{equation}
for the non-chiral theory and
\begin{equation}
 a_l = \frac{5\, g^2_M}{96 \pi^2} \left( \frac{m_l}{M} \right)^2  \ ,
\label{eq:g2_chiral}
\end{equation}
where $l=(e,\mu,\tau)$, if the particle in the multiplets are left-handed.
In (\ref{eq:g2_nonchiral}) and (\ref{eq:g2_chiral}) only
the leading contributions in $m^2_l / M^2$, where $m_l$ is the lepton mass and $M$ the \mulato mass, are
taken into account.

The particle data group \cite{pdg2010} quotes the following values for the anomalous magnetic moment
\begin{displaymath}
  a_l = \left( \frac{g-2}{2} \right)_l = \left\{
  \begin{array}{lll}
   \left( 1159.65218073 \pm 0.00000028 \right) \times 10^{-6} & & \mbox{ for } l = e \, ,  \\
   & & \\
   \left( 11659208.9 \pm 5.4 \pm 3.3 \right) \times 10^{-10} & & \mbox{ for } l = \mu \, , \\
   & & \\
   >  -0.052 \mbox{ and } < 0.013  & & \mbox{ for } l = \tau \, .
  \end{array} \right.
\end{displaymath}
Further, for the muon there is a $3.2\,  \sigma$ difference between the experimental value $a^{exp}_\mu$
and the standard model prediction $a^{SM}_\mu$ which is of
\begin{displaymath}
   \Delta a_\mu = a^{exp}_\mu  - a^{SM}_\mu = 255 (63) (49) \times 10^{-11} \, .
\end{displaymath}

For the non-chiral theory the \mulato contribution to the lepton anomalous magnetic moment is, for the electron
and for the $\mu$, negative due to the $\tau$ loop correction to the vertex. Therefore, in the non-chiral theory, the
\mulato cannot explain $\Delta a_\mu$ and $a_{e, \mu}$ should be, at most, of the order of the experimental error.
This provides the constraints
\begin{equation}
    \frac{g^2_M}{M^2} \leq 6.50 \times 10^{-14} \mbox{ MeV}^{-2} \quad \mbox{ or } \quad
    \frac{g^2_M}{M^2} \leq 8.14 \times 10^{-13} \mbox{ MeV}^{-2}
\end{equation}
if one uses $a_e$ or $a_\mu$; for $a_\mu$ the errors reported in \cite{pdg2010} were added in quadrature.
In the above calculation we used $m_e = 0.511$ MeV and $m_\mu = 105.658$ MeV.
This bounds can be rewritten in terms of the \mulato mass as
\begin{equation}
    M \geq g_M \times 3.9 \mbox{ TeV} \quad \mbox{ and } \quad
    M \geq g_M \times 1.1 \mbox{ TeV},
\end{equation}
respectively. The \mulato contribution to the $\tau$ anomalous magnetic momenta should be $ \sim 1.5 \times 10^{-9}$ or smaller.

For the chiral, the  \mulato contribution to $(g-2)/2$ should comply with the above results and can be, at most, of the order
of the muon anomaly $\Delta a_\mu$, i.e.
$ a_\mu \leq 255 (63) (49) \times 10^{-11} $, therefore
\begin{equation}
   \frac{g^2_M}{M^2} \leq 4.33 \times 10^{-11} \mbox{ MeV}^{-2}
\end{equation}
and the \mulato mass should
\begin{equation}
   M  \geq g_M \times 0.152 \mbox{ TeV} \, .
\end{equation}

For the non-chiral theory, the choice of $\Delta a_\mu$
to define the \mulato mass complies with the experimental error for the electron
and tau. Indeed, from (\ref{eq:g2_chiral}) it follows that the contribution of
the new gauge bosons to the electron/tau magnetic moment is
\begin{equation}
  a_l = \frac{m^2_l}{m^2_\mu} \, a_\mu \, .
\end{equation}
These scaling laws give an $a_e = 6.0 \times 10^{-14}$  and $a_\tau = 7.2 \times 10^{-7}$
which are smaller than the experimental error.

We call the reader attention that we are assuming a perturbative solution for the theory, i.e. that $g_M \ll 1$,
and the bounds derived from the magnetic moment for the \mulato mass can be of the same order of magnitude as
the electroweak scale.


The Lagrangian (\ref{L_q_mulato}) allows for flavor changing processes within the same family.
The \mulato propagator is flavor diagonal, therefore only those processes where the propagator links
the same type of vertices at both ends can have a non-vanishing $S-$matrix. The following lepton family
number violating decays
\begin{eqnarray}
&&\mu^- \longrightarrow
e^- \nu_e \overline\nu_\mu , \ \
\mu^- \longrightarrow e^- e^+ e^- , \ \
\tau^- \longrightarrow e^- e^+ e^-, \nonumber
\\
&&\tau^- \longrightarrow  e^- \mu^+ \mu^-, \ \
\tau^-\longrightarrow \mu^- e^+ e^-, \ \
\tau^- \longrightarrow \mu^+ e^+ e^-,  \nonumber
\end{eqnarray}
can only occur if the vertices connected by the \mulato propagator are different and, therefore, they
are forbidden within the model. On the other hand, the quark - \mulato vertex structure gives a
vanishing $S-$matrix for $\mu^- \rightarrow e^- \nu_e \overline \nu_\mu$. Processes with photons,
such as, $\mu^- \rightarrow e^- \gamma$ or $\mu^- \rightarrow e^- \gamma\gamma$ can only occur via loops
and are highly suppressed at low energies. The same arguments apply to process
$B_d \rightarrow e^-\tau^+$ and $B^0_s \rightarrow \mu^+\tau^-$ which are forbidden in the model.
It turns out that the gauge theory described here complies with the lepton flavor violation bounds
reported in the particle data book.


\begin{figure}[h]
\centering
\epsfig{file=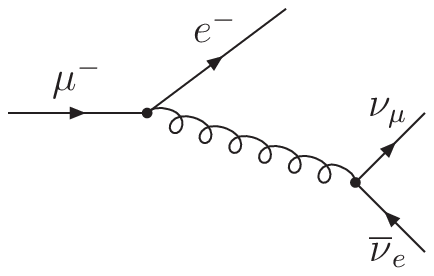,angle=0, width=6cm}
\vspace{.5cm}
\caption{Muon decay to $(e^-,\nu_\mu,\overline \nu_e)$ by \mulato exchange process.} \label{mudecay}
\end{figure}

The \mulato can also contribute to the leptonic decays of the $\mu$, the $D$'s and the $B$'s mesons. Let us now
discuss the bounds coming from this processes.
We start by computing the main muonic decay channel $\mu^- \rightarrow e^- \nu_\mu \overline\nu_e$ as
shown in fig. \ref{mudecay}. The $S-$matrix
gets a contribution from $W$ exchange and \mulato exchange. To leading order in $M_W$ and $M$, for the
chiral theory the matrix element for the transition reads
\begin{equation}
  \overline{ \left(  i \mathcal{M} \right)^2 } = 64 \, G^4_F \, \left[ 1 - \frac{1}{2 \sqrt{2}}
\frac{ g^2_M /M^2}{G_F} \right] \,
  \left(p_\mu \cdot p_{\overline\nu_e} \right) \,
  \left(p_e \cdot p_{\nu_\mu} \right) \, .
\end{equation}
For the chiral theory an extra factor of $1/2$ should multiply the $g^2_M/M^2$.
The \mulato contribution can be viewed as a modification to the Fermi coupling constant, i.e.
\begin{equation}
   G_F \longrightarrow G_F \, \left[ 1 - \frac{1}{2 \sqrt{2}} \frac{ g^2_M /M^2}{G_F} \right]^{1/4}  \approx G_F \, \left[ 1 - \frac{1}{8 \sqrt{2}} \frac{ g^2_M /M^2}{G_F} \right]  \, ,
\end{equation}
with \cite{pdg2010}, $G_F = 1.16637(1) \times 10^{-5}$ GeV$^{-2}$.
Requiring that the \mulato contribution to be of order of
the error on $G_F$ or smaller, gives the following bound
\begin{equation}
  \frac{1}{8 \sqrt{2}} \frac{ g^2_M}{ M^2}  \le 1.0 \times 10^{-10} \, \mbox{ GeV}^{-2} \qquad \mbox{ or } \qquad
  \frac{ g^2_M}{ M^2}  \le 1.13 \times 10^{-9} \, \mbox{ GeV}^{-2} \, .
  \label{bound_tramado}
\end{equation}
If instead of the Fermi coupling constant, one uses the relative error on the
muon width $\Gamma_\mu = 4.799980(46) \times 10^{-17}$ MeV, the bound becomes
\begin{equation}
  \frac{1}{8 \sqrt{2}} \frac{ g^2_M}{ G_F \, M^2}  \le 9.58 \times 10^{-6}  \qquad \mbox{ or } \qquad
  \frac{ g^2_M}{ M^2}  \le 1.26 \times 10^{-9} \, \mbox{ GeV}^{-2} \, .
\end{equation}
The corresponding mass bounds are, respectively,
\begin{equation}
   M \geq  g_M \times 30 \mbox{ TeV} \qquad \mbox{ and } \qquad g_M \times 28 \mbox{ TeV} \, .
\end{equation}
Assuming that $g_M \approx e = \sqrt{4 \pi \alpha} = 0.30$, we have a lower bound the \mulato mass
of $\approx 9$ TeV to comply with both the $\beta$-decay, i.e. the error on the Fermi coupling constant,
and the muon decay. These mass bounds are more restrictive than the bounds coming from the
anomalous magnetic moment.


The  \mulato vertices can give rise to the following leptonic decays
\begin{displaymath}
 D^0 (c \overline u) \longrightarrow \mu^- e^+, \quad
 B^0 (b \overline d) \longrightarrow \tau^- e^+, \quad
 B^0_s (s \overline b) \longrightarrow \tau^- \mu^+
\end{displaymath}
an its complex conjugate decays. The processes width can be computed using the relation
\begin{equation}
   \langle 0 | ~ \overline q \, \gamma^\mu \gamma_5 \, q^\prime ~ | (q \overline q^\prime ) \rangle = i \, f \, P^\mu \, ,
\end{equation}
where $| (q \overline q^\prime ) \rangle$ stands for the meson state composed of
quarks $q \overline q^\prime$, $f$ is the meson decay constant and $P$ the four-momentum of the meson.
Note that the above decays are possible only within the chiral theory. The heavy meson leptonic decay width
is calculated by evaluating the amplitude shown in fig. \ref{ddecay}, exemplified for $D^0\to\mu^-e^+$, which
in the general case is given by:
\begin{equation}
  \Gamma = \frac{1}{256 \, \pi} \, \frac{g^4_M}{M^4} \, f^2 \, m^2_l \, m_m \, \left( 1 - \frac{m^2_l}{m^2_m} \right)^2 \, ,
  \label{largura_mesao}
\end{equation}
where we have assumed that the lightest lepton is massless, $m_l$ is the mass of the heavier lepton and
$m_m$ the mass of the meson state.

For the decay $D^0 \rightarrow \mu^- e^+$, using a $m_{D^0} = 1.864$ GeV and $f_{D^0} = 0.206$ GeV, from the bound
(\ref{bound_tramado}) if follows that the corresponding branching ratio should satisfy
\begin{equation}
  Br(D^0 \rightarrow \mu^- e^+) < 8.7 \times 10^{-13} \,
   \label{bound_D0}
\end{equation}
to be compared with the experimental limit \cite{pdg2010} of
\begin{equation}
  Br(D^0 \rightarrow \mu^- e^+) < 2.6 \times 10^{-7}\, .
\end{equation}

For the decay $B^0 \rightarrow \tau^- e^+$, using a $m_{B^0} = 5.279$ GeV \cite{pdg2010}
and $f_{B^0} = 0.22$ GeV \cite{Mohanta2010}, the bound (\ref{bound_tramado}) gives a
\begin{equation}
  Br(B^0 \rightarrow \tau^- e^+) < 2.3 \times 10^{-9}
   \label{bound_B0}
\end{equation}
to be compared with the experimental limit \cite{pdg2010} of
\begin{equation}
  Br(B^0 \rightarrow \tau^- e^+) < 2.8 \times 10^{-5}\, .
\end{equation}

Finally, for the decay $B^0_s \rightarrow \tau^- \mu^+$, using a $m_{B^0_s} = 5.366$ GeV \cite{pdg2010}
and $f_{B^0_s} = 0.24$ GeV \cite{Mohanta2010}, the bound
(\ref{bound_tramado}) gives a
\begin{equation}
  Br(B^0_s \rightarrow \tau^- \mu^+) < 2.7 \times 10^{-9}  \, .
  \label{bound_B0s}
\end{equation}
To the best knowledge of the authors, for this decay there is no experimental information.

The $D^0$ and $B^0$ branching ratios are at least two orders of magnitude smaller than the experimental upper bounds.
According to our estimates, there branching ratios should be quite smaller. For $B^0_s$, the
experimental bounds coming from $g-2$ and muon decay predicts a branching ratio of the same order of magnitude as
for $B^0$. Note that, in the standard model, the decays discussed are not allowed at tree level but they are allowed
if one consider one-loops diagrams.

The same type of processes can give rise to the production of dark matter.
For example, the bounds (\ref{bound_D0}),  (\ref{bound_B0}), (\ref{bound_B0s}) also apply
to the decays where the leptons are replaced by their dark counter parts. These bounds suggests
that the branching rates for production of dark matter from $D$, $B^0$ and $B^0_s$ decays are, at most,
of the order of  $10^{-9}$. Note that the width are proportional to the lepton mass squared and vanish
for massless particles in the final state.


Our estimates for the \mulato mass suggest a $M \geq 9$ TeV. Then, from the point of view of the \mulato all
the particles in the multiplets (\ref{matter}) are massless. This simplifies considerably the computation of
the \mulato width. It follows that,
\begin{equation}
   \Gamma = \frac{g^2_M \, M}{24 \pi} \, N_F \, ,
   \label{gamma_1_8}
\end{equation}
where $N_F$ is the number of multiplets.
For the chiral theory (\ref{gamma_1_8}) should be multiplied by $1/2$. The bound (\ref{bound_tramado})
gives
\begin{equation}
   \Gamma \leq 1.5 \times 10^{-5} \, M^3 \, N_F,
\end{equation}
respectively. $\Gamma$ and $M$ are given in TeV. It follows that $\Gamma \approx 0.8$ TeV or smaller and, therefore,
the \mulato should have a very short lifetime $\tau = 1 / \Gamma \approx 8 \times 10^{-28}$ s.
This means that in the cosmic rays either the \mulato is produced via an
high energy process or it is absent from the cosmic rays spectrum.



The \mulato interaction can give rise to processes which are forbidden in the Standard Model. In particular for the
collision  $e^+ + p \rightarrow \mu^+ + X$ where $X = \Lambda$ or $\Lambda_c$ can occur via $t$-channel
\mulato exchange but is forbidden at tree level in the Standard Model. At the parton level, the tree-level
amplitude for lepton-quark scattering with violation of the lepton family number and flavor exchange is shown in
fig. \ref{flavorx} (a) and (b), for charm and strangeness production, respectively. Note that processes
$e^- + p \rightarrow \mu^- + X$ is forbidden at tree-level within our model.

The differential cross section at the parton level is
\begin{equation}
 \frac{d \sigma}{d \Omega} = \frac{1}{1024 \, \pi^2} \, \frac{g^4_M}{M^4} \, s \, \left(1 - \frac{m^2}{s} \right)^2
 ( 1 + \cos\theta ) \left[ 2 - \left(1 - \frac{m^2}{s} \right)( 1 - \cos\theta ) \right] \,
\end{equation}
where $s$ is the c.m energy, $m$ the mass of the quark in the final state and $\theta$ the angle between the $\mu^+$
and $e^+$ momentum. The total cross section reads
\begin{equation}
 \sigma (s) = \frac{1}{384 \, \pi} \, \frac{g^4_M}{M^4} \, s \, \left(1 - \frac{m^2}{s} \right)^2
 \, \left(2 + \frac{m^2}{s} \right) \, .
\end{equation}
From the bound (\ref{bound_tramado}), it follows that
\begin{equation}
 \sigma (s) \leq 8.2 \times 10^{-13} \left( \frac{s}{1 \, \mbox{GeV}^2} \right) ~ \mbox{pbarn} \, .
\end{equation}


So far, we have investigated the visible sector to constrain the parameters of
the model through comparison with well established experimental results. The data clearly provide
acceptable bounds for $g_M/M$ and set a scale for the \mulato lifetime. 
Experimental limits for the WIMP-nucleon interaction cross section, from 
precision recoil measurements  on different nuclear targets, have been reported recently
\cite{CDMS,XENON100,XENON100-2011,DAMA,DAMA2010,DAMA2010b,COGENT}.
Our \mulato model allows to compute this type of process involving dark and nuclear particles by
constructing  the expected Fermi-like point interaction from our basic lagrangean (\ref{lagrangeano}).
This work is in progress and will be reported elsewhere \cite{futuro}.


In summary, we have deeveloped a detailed model for the particle interactions between
Dark Matter and Visible Matter. We go beyond the Standard Model by postulating the existence of six
parallel universes, one of which is our visible universe which is composed of two charge families of
quarks, u, c, and t of charge 2e/3; d, s and b, of charge -e/3; the leptons and their corresponding
neutrinos. The other five universes contain only DM in  similar two charge families of dark quarks,
$u_d, c_d, t_d$ of charge 2e/3, $d_d, s_d, b_d$, of charge -e/3; leptons, $e_d$, $\mu_d$ and $\tau_d$,
and dark neutrinos.  The boson responsible for the interactions between the DM particles and the VM ones,
the \mulato, is estimated to have a mass much larger than the electroweak mass scale. We stress that the model
is economic in the number of parameters and the phenomenological implications for fermionic processes at
tree-level require only the new gauge coupling and \mulato mass. Several decay modes
and other processes are calculated and upper bounds are established for them

The existence of the interaction associated with a new gauge boson, the \mulato with a mass of the order
of TeV or higher, implies that a new phase transition should happen in early Universe before the
electroweak one. The new phase transition corresponds to the scale where the visible and dark matter
decouple. Then, for lower temperatures, dark and visible matter see each other mainly through gravity.
The implications and signatures of this new phase transition remain to be investigated.

\section*{Acknowledgements}

The authors acknowledge financial support from the Brazilian
agencies FAPESP (Funda\c c\~ao de Amparo \`a Pesquisa do Estado de
S\~ao Paulo) and CNPq (Conselho Nacional de Desenvolvimento
Cient\'ifico e Tecnol\'ogico) and the US Department of Energy
Grants DE-FG02-08ER41533, DE-SC0004971, DE- FC02-07ER41457 (UNEDF, SciDAC-2) and
the Research Corporation. OO acknowledges financial support from FCT under
contract PTDC/\-FIS/100968/2008.


\end{document}